\documentclass{ws}

\def \bra {\langle}
\def \ket {\rangle}
\def \barr {\begin{array}{lc}}
\def \earr {\end{array}}
\def \comma {\; \; , \;}
\def \up  {| \! \uparrow \ket}
\def \down {| \! \downarrow \ket}
\def \ud  {| \! \uparrow \downarrow \ket}
\def \lf  {\left (}
\def \rt  {\right )}
\def \d   {\delta}
\def \eps {\epsilon}
\def \sg  {\sigma}
\def \gm  {\gamma}
\def \Gm  {\Gamma}
\def \ZZ  {\mathbb{Z}_2}

\begin{document}

\begin{center}

{\hbox to \hsize{\hfill CU-TP-1106}}

{\Large \bf De Sitter Space With Finitely Many States: \\
A Toy Story}

\bigskip
\bigskip
\bigskip

{\large \sc Maulik~Parikh\footnote{\tt mkp@phys.columbia.edu}
 {\rm and}
  Erik~Verlinde\footnote{\tt erikv@science.uva.nl}\\[1cm]}

{\it $^a$ Department of Physics, Columbia University, New York, NY
  10027}\\[3mm]

{\it $^b$ Physics Department, University of Amsterdam, The Netherlands}\\[3mm]

\vspace*{1.5cm}

\large{\bf Abstract}
\end{center}
The finite entropy of de Sitter space suggests that in a theory of 
quantum gravity there are only finitely many states. It has been argued 
that in this case there is no action of the de Sitter group  
consistent with unitarity. In this note we propose a way out of this
if we give up the requirement of having a hermitian
Hamiltonian. We argue that some of the generators of the de
Sitter group act in a novel way, namely by mixing in- and out-states. 
In this way it is possible to have a unitary S-matrix that is
finite-dimensional and, moreover, de Sitter-invariant. 
Using Dirac spinors, we construct a simple toy
model that exhibits these features.\footnote{Based in part on a talk
  presented by one of us (M.~P.) at the Xth Marcel Grossmann Meeting,
  held in Rio de Janeiro, July 20-26, 2003.}

\noindent

\bigskip

\section{Introduction}

In AdS/CFT, the $O(2,4)$ isometry group of anti-de Sitter space is
reinterpreted as conformal symmetry of the dual theory. For de Sitter
space, we do not yet possess the holographically dual theory;
nevertheless, we would expect the $O(1,d)$ isometry group of de Sitter
space to be a symmetry of that theory.
But, unlike AdS, de Sitter space also has horizons, with
an associated entropy. There are differing interpretations of the
entropy but one view, which we shall explore here, is that it
indicates that the dual theory has only a finite number of
states\cite{tombanks,fischler}.

However, there is a well-known theorem that says that there are no
nontrivial finite-dimensional unitary representations of a noncompact
group. Consequently, the Fock space of the dual theory can be either
finite-dimensional or a unitary representation of the de Sitter group, but
not both. There have thus been claims that de Sitter space has no
holographic dual\cite{gks}, or that the symmetry group is not the de
Sitter group\cite{Witten,alberto}.

Here we will consider the possibility that the dual to de
Sitter space is indeed defined by a finite-dimensional Fock space, so
that the entropy is finite. The finiteness of the Fock space requires
that particles in the dual theory obey Fermi-Dirac statistics. Hence we
shall construct a spinor dual to de Sitter space. The
finitely many Fock states describing a static patch of de Sitter space
do not form representations of the de Sitter group but only of the (compact)
rotation subgroup. Global de Sitter space is described by two copies
of the Fock space. After antipodal identification, these become the
space of initial and final states. Some of the de Sitter generators
then mix the in- and out-states i.e. bras with kets. The main idea is
that there are certain pairings of bras and kets that are nevertheless
de Sitter-invariant. These can be used to construct a de Sitter-invariant
S-matrix so, in that sense, the isometry group of de Sitter space
indeed appears as a symmetry of the dual theory. In this paper, we
will illustrate the reconciliation of symmetry and entropy with a toy
model based on Dirac spinors; a more refined construction involving
fuzzy spheres will appear in a companion paper\cite{finitereps}.

\section{Fock States and Tensor Product States}
De Sitter space is conveniently thought of as a timelike hyperboloid embedded
in Minkowski space:
\begin{equation}
-{X^0}^2 + {X^1}^2 \dots + {X^d}^2 = +R^2 \; , \label{embed}
\end{equation}
where $X^I$ are Cartesian embedding coordinates. In this form, the
$O(1,d)$ isometry group of $d$-dimensional de Sitter space is
manifest: it is
the Lorentz group in $d+1$ dimensions.
Now, consider a geodesic observer in de Sitter space.
Without loss of generality, let
the observer be at the north pole, which we take to be in the
positive $X^d$ direction.
The Lorentz generators that leave the observer's worldline invariant are the
rotations about the axis connecting the poles as well as the boost in
the $X^d$ direction.
Together these form the observer's little group,
$O(d-1) \times R$. Note that since the rotation group $O(d-1)$ is
compact, it admits finite-dimensional unitary representations.

Now, since the states counted by the entropy are associated with a
horizon, we should take the Hilbert space to form a representation of
the group that keeps the horizon fixed. This is just the observer's little
group. Hence we take the one-particle Hilbert space,
$h_I$, to be a finite-dimensional unitary representation of
$O(d-1)$. (It is, however, a nonunitary representation of $O(d-1)
\times R$, as we will see.) We
denote the corresponding multi-particle Hilbert space, or Fock space,
by $H_I$. In order to obtain a finite-dimensional Fock space, the particles
need to obey Fermi-Dirac statistics.
Similarly, the
states accessible to the antipodal observer at the south pole
belong to a finite-dimensional Fock space, $H_{II}$, constructed out of a
one-particle Hilbert space, $h_{II}$, that is isomorphic to that of
the observer at the north pole.
The state describing global de Sitter space
is a tensor product of states in the Fock spaces of antipodal pairs of
observers. A typical basis state of the dual to global
de Sitter space is therefore
\begin{equation}
|m \ket_I \otimes |n \ket_{II} \; ,
\end{equation}
where $m$ and $n$ are
quantum numbers of
$O(d-1)$. The general state is
\begin{equation}
|\Psi \ket = \sum_{m, n} C_{mn} |m \ket_I \otimes |n \ket_{II} \; . \label{psi}
\end{equation}
Mixed states that cannot be written in the form
$|\psi \ket_I \otimes |\phi \ket_{II}$ are those
that correlate the antipodal pair of observers. The state
corresponding to a particle beyond the future horizons of both the
observers would be an example of a mixed state.

The tensor product states live in the direct product space $H_I
\otimes H_{II}$. This implies that the one-particle Hilbert
space for global de Sitter space is a direct sum of the two observers'
Hilbert spaces. One can check this by counting the number of states:
if there are $n$ states in an observer's one-particle Hilbert space, then
there are $2^n$ states in the Fock space, assuming Fermi-Dirac
statistics, and hence $2^n \times 2^n =
2^{2n}$ states in the global (tensor product) Fock space. So the global
Hilbert space, $h$, has $2n$ states,
indicating that it is a direct sum of the two Hilbert spaces:
\begin{equation}
h = h_I \oplus h_{II} \; .
\end{equation}

Consider now a general de Sitter transformation, $M$. The Fock spaces
$H_I$ and $H_{II}$ are $O(d-1)$-invariant by
construction, but they are not de Sitter-invariant. This is because a
de Sitter transformation that is not in $O(d-1) \times R$
can take a particle out of an observer's horizon into
the antipodal observer's horizon. So on the global one-particle
Hilbert space, $h$, $M$ acts as $M: h \to h$ i.e. as
\begin{equation}
M: h_I \oplus h_{II} \to h_I \oplus h_{II} \; . \label{dSonHilbert}
\end{equation}
More generally, on the
space of tensor products of Fock states, $M$ acts as
\begin{equation}
M: H_I \otimes   H_{II} \to   H_I \otimes  H_{II} \; .
\end{equation}
Of course, since the individual Fock spaces are finite-dimensional,
the tensor product space is also finite-dimensional. Hence it also
falls afoul of the theorem forbidding nontrivial unitary finite-dimensional
representations of the de Sitter group.
But we can
find certain special tensor product states that transform in a {\em
trivial} representation of the de Sitter group. That is, we will
look for those tensor product states, $|S \ket$, that are
invariant under the de Sitter group, $M|S \ket = 0$.
The Fock space accessible to an individual observer nevertheless has a
finite number of states, so the entropy is finite.
We will argue that if we identify antipodal points in de Sitter space, the
tensor product states become
processes; the special
tensor product states, $|S \ket$, then correspond to de
Sitter-invariant processes.

\section{A Toy Model for de Sitter Space}

We now construct a toy model based on Dirac spinors. This is perhaps
the simplest dual to de Sitter space. There are a couple of
motivations for considering spinors. Because of spin-statistics, the
Fock space of fermions can be finite-dimensional. For
bosons, the Fock space would be infinite-dimensional even for a
finite-dimensional Hilbert space as there would be no restriction on the
number of bosons. Another reason for considering spinor
representations is that the number of components of a Dirac spinor doubles
when the dimension is increased by two. So for Dirac representations,
$R$, we have
\begin{equation}
{\rm dim}~ O(d-1)_R = {1 \over 2} {\rm dim}~ O(1,d)_R \; .
\end{equation}
This allows us to write a global Hilbert state, which transforms under
$O(1,d)$, as a direct sum of two states that transform under $O(d-1)$,
in accordance with (\ref{dSonHilbert}).

The gamma matrices obey the Clifford algebra $\{\Gm_I, \Gm_J \} = 2
\eta_{IJ}$, where $I, J = 0 \ldots d$.
A convenient representation is
\begin{equation}
\Gm_i = \sg_3 \otimes \gm_i = \lf \barr \gm_i \; \; \; \; 0 \\ 0 \;
-\gm_i \earr \rt \comma \Gm_0 = i \sg_2 \otimes 1 = \lf \barr \; \; 0
\; \; 1 \\
-1 \; 0 \earr \rt \comma \Gm_d = \sg_1 \otimes 1 = \lf \barr 0 \; \; 1 \\ 1
\; \; 0 \earr \rt \; ,
\end{equation}
where $\gm_i$ are the gamma matrices for the Clifford algebra
$\{\gm_i, \gm_j\} = 2 \d_{ij}$, with $i, j$ running from 1 to
$d-1$. The top half of the $\Gamma$ matrices act on a spinor state in
$h_I$, with the bottom half acting on $h_{II}$.

The de Sitter generators can then be written as
\begin{equation}
M_{IJ} =  -{i \over 4} [ \Gm_I , \Gm_J ] \; .
\end{equation}
It is convenient to express
the de Sitter generators in terms of indices $i, j$ that run from 1 to $d-1$:
\begin{equation}
J_{ij} \equiv M_{ij} \qquad P_i \equiv M_{di} \qquad K_i \equiv M_{0i}
\qquad H \equiv M_{0d} \; .
\end{equation}
Here $J_{ij}$ generate the rotation group $SO(d-1)$, $P_i$ are
momentum operators, $K_i$ are the boosts, and $H$ is the generator of
time translations.
To be specific, consider four-dimensional de Sitter space. The
four-dimensional de Sitter group is $O(1,4)$ and the observer's
little group is $O(3)$. Then, in
the above representation, the de Sitter generators are
\begin{equation}
J_i = {1 \over 2} \lf \barr
\sg_i \; \; 0 \\ 0 \; \; \sg_i \earr \rt \comma
P_i = {i \over 2} \lf \barr \; \; 0 \; \; \sg_i \\ -\sg_i \; 0 \earr \rt \comma
K_i = {i \over 2} \lf \barr 0 \; \; \sg_i \\ \sg_i \; \; 0 \earr \rt \comma
H = {i \over 2} \lf \barr -1 \; 0 \\ \; \; 0 \; \; 1 \earr \rt \; ,
\end{equation}
where $\sg_i$ are the Pauli matrices and $J_i = {1 \over 2} \eps_{ijk}
J_{jk}$. Notice that $P_i$ and $K_i$ are off-diagonal, indicating that
they do not act within the Hilbert space of a single observer.
In this representation, $J_i$ and $P_i$ are hermitian whereas $K_i$
and $H$ are not. That the Hamiltonian, $H$, is anti-hermitian may
seem odd, but it may just indicate that any excitation of empty de Sitter
space decays to empty de Sitter space. It should also be stressed that
$H$ acts on states in quantum gravity which presumably
live at the boundary; there is no obvious connection between the
eigenvalues of $H$ and the energy measured perturbatively by an
observer in the bulk.

The one-particle Hilbert space, $h_I$, consists of just two
states: the two-component spinors, $\up$ and $\down$. The antipodal
observer has an isomorphic Hilbert space, $h_{II}$, and the above
$4 \times 4$ matrices act on the direct sum of these two Hilbert
spaces. The Fock space consists
of the four states $|0 \ket$, $\up$, $\down$, and $\ud$. These four
states can be labeled by their transformation properties under
$O(3)$. Obviously, they belong to the $\bf{1}$, the $\bf{2}$, and
the $\bf{1}$ representations, respectively.

The antipodal observer has an isomorphic Fock space. The tensor
product states therefore transform under the direct product of the
respective representations. Taking the direct product of the $\bf{1}$,
the $\bf{2}$, and the $\bf{1}$ with themselves we find that the 16 tensor
product states belong to
\begin{equation}
{\bf 1} \oplus {\bf 2} \oplus {\bf 1} \oplus
{\bf 2} \oplus {\bf 3} \oplus {\bf 1} \oplus {\bf 2} \oplus
{\bf 1} \oplus {\bf 2} \oplus {\bf 1} \; .
\end{equation}
These are $O(3)$ labels. We are interested in how the
tensor product states transform not just under rotations, but under
general de Sitter transformations. We can group the above
$O(3)$ representations into $O(1,4)$ representations:
\begin{equation}
{\bf 1} \oplus {\bf 4} \oplus {\bf 5} \oplus
{\bf 1} \oplus {\bf 4} \oplus {\bf 1} \; . \label{so14reps}
\end{equation}
For example, the states $|0\ket_I \otimes \up_{II}$,
$|0\ket_I \otimes \down_{II}$,
$\up_I \otimes | 0 \ket_{II}$, and
$\down \otimes |0 \ket_{II}$ together combine to form a $\bf{4}$,
which is a spinor of $O(1,4)$. The $\bf{5}$ is a vector of $O(1,4)$.

Notice, in particular, that there are three singlet states. These are
tensor product states that are invariant under the action of the de
Sitter group. The three invariant states are
\begin{equation}
|0\ket \otimes |0 \ket \comma \ud \otimes \ud \comma \lf
\up \otimes \down - \down \otimes \up \rt \; .
\label{invariants}
\end{equation}

\section{A De Sitter-Invariant S-Matrix}

So far we have considered the Fock spaces $H_I$ and $H_{II}$ as being
isomorphic but otherwise independent spaces. De Sitter-invariance is
implemented through choosing particular tensor products of states that
are invariant under the full de Sitter group. This group-theoretic
construction is correct as far as it goes but we can make it more
compelling from a physical standpoint if we consider
antipodally-identified or ``elliptic'' de Sitter
space\cite{Schroedinger,dSZ2,Gibbons}.
This consists of identifying
events that are related by the $\ZZ$ antipodal map
\begin{equation}
X^I\to -X^I \; ,
\end{equation}
where $X^I$ are the Cartesian embedding coordinates, together with
charge conjugation, C\cite{dSZ2}.

Antipodal identification identifies points at time $X^0$ with
those at $-X^0$. As a result, elliptic de Sitter space is not
time-orientable; there is no globally-consistent way to distinguish
future light cones from past light cones.
In particular, the arrow of time in the antipodal observer's causal
patch points in the opposite sense. This suggests that if we think of $H_I$
as the space of initial kets, then the antipodal Fock space should be
regarded as the space of {\em final bras}\cite{dSZ2,gerard}. The tensor
product states $|m \ket_I \otimes | n\ket_{II}$ can now be thought
of as a physical process $|m \ket_{\rm in} \to |n' \ket_{\rm out}$.
Here the $'$ indicates that the out-state $| n' \ket_{\rm out}$ is 
actually a charge, time and parity reversal of the corresponding
$|n \ket_{II}$ e.g. $\up' = \down$. We also have a choice, whether to
map $|0 \ket_{\rm in}$ to $\bra 0|_{\rm out}$ or to $\bra \,
\uparrow \downarrow \! |_{\rm out}$. Then taking the hermitian conjugate of
(\ref{psi}), we find
\begin{equation}
\Psi = \sum_{m,n} C^*_{mn} | n' \ket_{\rm out} \otimes 
\bra m|_{\rm in} \; , 
\end{equation}
and hence $C^\dagger$ is just the S-matrix. The invariant tensor
product states (\ref{invariants}) are then the de Sitter-invariant
building blocks of the S-matrix. For example, with $|0 \ket_{\rm in}
\to \bra \, \uparrow \downarrow \! |_{\rm out}$, and with the rows and
columns labeled in the order $|0 \ket$, $\up$, $\down$, $\ud$, a de
Sitter-invariant S-matrix is
\begin{equation}
S = \lf \barr 
0 \; \; \; 0 \; \; \; 0 \; \; \, a \\
0 \; \; \; b \; \; \; 0 \; \; \; 0 \\
0 \; \; \; 0 \, \; \; b \; \; \; 0 \\
c \; \; \; 0 \; \; \; 0 \; \; \; 0 
\earr \rt \; ,
\end{equation}
where we have used $T \up = \eta \down$ and $T \down = - \eta \up$
for time-reversal on spinors. If $a$, $b$, and $c$ are all phases,
then the S-matrix is unitary as well as de Sitter-invariant.

We have shown by construction that a finite-dimensional Fock space is
not in conflict with the de Sitter symmetries. To illustrate the point, we
used a very small representation in which there were as few as four
states. We mention, though, that by considering larger
representations, one can see the emergence of geometric structure in
the form of a fuzzy sphere at the boundary of de Sitter
space\cite{finitereps}.

M.~P. is supported by DOE grant DF-FCO2-94ER40818.

\end{document}